# A House Monotone, Coherent, and Droop Proportional Ranked Candidate Voting Method

Ross Hyman
rossahyman@gmail.com

**Abstract**
A proportional ranked candidate voting method is described that satisfies ranked candidate analogs of the voting criteria satisfied by Phragmén's approval voting procedure. The method complies with Droop proportionality, house monotonicity, and coherence. The last two properties are the ranked-candidate analogs of the divisor method properties of always avoiding the Alabama and New State paradoxes. The method is free rider resistant and adding a candidate to each ballot that is universally preferred to all other candidates does not alter the relative winning order of the other candidates. The highest ranked candidate in the list is the Instant Runoff winner, which is shown to be in at least one Droop proportional set of N winners for all N, giving a quantitative meaning to the notion that the Instant Runoff winner has both core and broad voter support.

## Section I: Introduction

The United States Constitution requires that the number of congressional representatives apportioned to each state be proportional to the population of the state. From 1850 to 1900, apportionments were determined by a method described by Alexander Hamilton. Two flaws were found in Hamilton's method in the late nineteenth and early twentieth centuries. The first flaw, called the Alabama Paradox, was discovered in 1880 when the number of seats apportioned to Alabama decreased if the total number of seats increased. The second flaw, called the New State Paradox, was discovered in 1907 when Oklahoma became a state. When Oklahoma was included in the apportionment calculation and the total number of seats were increased so that the total number of seats apportioned to states other than Oklahoma was the same number before Oklahoma was added, New York lost a seat and Maine gained a seat (Balinski, M and Young, H 2001).

Divisor methods are apportionment methods that do not exhibit these flaws. The U.S. Congress now apportions representatives using a divisor method (Balinski, M and Young, H 2001) and most countries that elect parliaments in proportional representation elections also use divisor methods to determine how many seats should be apportioned proportionally to each party's vote (Pukelsheim, F. 2017).

However, Single Transferable Vote (STV) (Tideman, Nicolaus 2006), a proportional representation election method used in Australia, Ireland, Scotland, and New Zealand, and increasingly in the U.S. is subject to the Alabama and New State Paradoxes. STV elections allow voters the flexibility of ranking candidates according to their preference order, instead of voting for a single party. They comply with Droop proportionality for solid coalitions, which Woodall described as the most important property of STV and the sine qua non for a fair election rule (Woodall, Douglas R. 1994). Droop proportionality is described in Section II.

The ranked candidate analog of not exhibiting the Alabama paradox is called house monotonicity and is the property that the winners of an N-winners election should also be winners of the N+1-winners election. STV is not compliant with house monotonicity. When an election method satisfies house monotonicity, the winners can be placed in a proportional list in which the top N candidates in the list are the winners of the N-winners election. Being able to produce a proportional list is valuable since political parties are required to produce a candidate list for party-list proportional representation elections. Even when a party list is not required, a house monotonic compliant winning set is valuable in other ways. House monotonic winners of the N-winners election who are also winners of the M-winners election for M<N, have *broad support* since voters are drawn towards them as the number of winning possibilities decease from N to M, indicating a more inclusive appeal than winners that do not satisfy house monotonicity. Likewise, house monotonic winners of the N-winners election, being also winners of the M-winners election for M>N, have a *core support* of voters who stick with them as the number of winning options increase. (We show that core support is also a property of STV even though it is not house monotonic). House monotonicity allows for the possibility of *free rider resistance* because ballots in an N winner election can be held accountable for contributing to the election of the N-1 previous winners even if those N-1 candidates are ranked lower on a ballot than the hopeful candidates contending for the N$^{\text{th}}$ winning position. This property is incompatible with the later-no-help/harm criteria complied with by STV. Finally, Balinski and Young (2001) showed that for an apportionment method to avoid the New State Paradox, it is a necessary but not sufficient condition for it to be house monotone.

The ranked-candidate analog of not exhibiting the New State Paradox is called coherence (Balinski 2005). If an N-winners election is applied to a ballot set that is composed of two sets of ballots, each completely ranking an entirely different set of candidates, and the N winners include M candidates from the first set and P candidates from the second set, then an M-winners election applied to the first set and a P-winners election applies to the second set should produce the same winners. STV is not compliant with coherence. If a method produces a proportional list, then coherence requires that the list produced by applying the method to either ballot set separately is the same list produced by excluding from the combined list the candidates ranked by the other ballot set.

In 1895, the mathematician Edvard Phragmén (Phragmén, E. 1895; 1899) described a divisor election method for approval style ballots that satisfies house monotonicity, coherence, and Droop proportionality. It also has the property that approving a new candidate on every ballot, places that candidate in the winning candidate list first, or amongst the other universally approved candidates, while preserving the entirety of the original winning candidate list below it.

In this paper, we extend Phragmén's divisor method to ranked-candidate ballots in a way that complies with Droop proportionality, house monotonicity, and coherence. It is free rider resistant and a candidate added first on every ballot will be first on the candidate list, while preserving the entirety of the original winning candidate list below it. We call this property *proportional candidate sharing*. Winners have both broad support and core support according to their placement on the winning candidate list. STV is also compliant with proportional candidate sharing and core support but this is not so for all Droop compliant methods.

In Section II, I define Droop proportionality. In Section III, I describe work by other researchers on devising ranked candidate methods that attempt to comply with house monotonicity and Droop proportionality. In Section IV, I describe work by other researchers on divisor-method-based ranked-candidates voting methods based on Phragmén's approval ballot method. In Section V, I describe a new top-down divisor-method-based ranked-candidates voting method based on Phragmén's method and in Section VI I demonstrate how it works with an example. In Section VII, I prove its compliance with house monotonicity, coherence, and Droop proportionality, and show that winners have both broad and core support. In Section VIII, I demonstrate free rider resistance and compliance with shared candidate proportionality. In Section IX, I summarize and conclude the paper.

**Section II: Droop Proportionality**

A **Droop quota** of ballots in an election with $V_{total}$ ballots to elect *N* winners is $Q_N = V_{total}/(N+1)$ ballots. The Droop quota is the smallest number of ballots such that if there are more than N candidates running for N seats and each of the N winners is supported by more than $Q_N$ ballots, then the losers must all have been supported by fewer than $Q_N$ ballots. For example, a candidate will have the support of more than half the voters to be the winner of a one-winner election, and all loosing candidates will have the support of fewer than half of the voters. Each of the two winning candidates will have the support of more than a third of the voters to be the winners of a two-winners election, and all the loosing candidates will have the support of fewer than a third of the voters, etc.

A collection of ballots containing the same *L* candidates above all others (not necessarily in the same order) is a **solid coalition** supporting those *L* candidates. For fully ranked ballots, which we will assume is the case for the proofs of compliance with Droop proportionality, every collection of ballots is a solid coalition, with a preferred set of candidates that can include, possibly, all candidates.

A set of the *L* candidates that are preferred over others on each ballot of a solid coalition is a **preferred candidate set** of that solid coalition. In general, a solid coalition can have several preferred candidate sets, smaller preferred candidate sets being subsets of the larger ones, with the set of all candidates being a preferred candidate set for all solid coalitions.

If more than *K* Droop quotas of voters form a solid coalition supporting *L* candidates, in which *K* is less than or equal to *L,* then **Droop proportionality** (Woodall, Douglas R. 1994) is satisfied if at least *K* of those *L* candidates are elected. Droop proportionality guarantees that if there are an odd number of seats, and there is a solid coalition of more than half the voters preferring a sufficient number of the candidates above all others, then more than half the seats will be filled from the set of those preferred candidates. This is so because if the solid coalition is greater than half the ballots, the number of guaranteed seats is (N+1)/2 rounded down. If N is odd, it can be written as N = 2(M+1/2) and the number of guaranteed seats is M+1, greater than half the seats.

A **Droop compliant set** for an election for *N* winners is a set of *N* candidates that satisfies Droop Proportionality.

For an example, consider the Example 1 ballot set. We use this example throughout the paper.

Example 1 ballot set: 60:A>D>C>B, 51:B>C>D>A, 45:C>A>D>B, 44:D>C>A>B.

There are 200 ballots. The Droop quota for one winner is 100 ballots. There is a solid coalition that ranks candidates A, C, and D above candidate B composed of 149 ballots, which exceeds one Droop quota for one winner. Therefore, the Droop compliant sets for the one-winner election are {A}, {C}, or {D}. The Droop quota for two winners is 66 2/3 ballots. The same 149 ballot solid coalition exceeds two Droop quotas for two winners. This means that {A, C}, {A, D}, and {C, D} are Droop compliant sets for two winners. The Droop quota for three winners is 50 ballots. There is a solid coalition placing candidate A above all other candidates, composed of 60 votes, and a solid coalition placing B above all candidates composed of 51 votes. The remaining ballots form a solid coalition supporting candidates A, C, and D above B, composed of 89 votes. Each of these solid coalitions exceeds one Droop quota for three winners. Because Droop compliance can be satisfied for the third set through the election of candidate A, which the first set requires to be elected, the third set does not add any further restrictions. The Droop compliant sets for three winners are {A, B, C} and {A, B, D}.

**Section III Attempts to Produce House Monotonic and Droop Proportion Lists**
Four house monotonic and Droop compliant candidate lists: A>C>B>D, A>D>B>C, C>A>B>D, and D>A>B>C, can be formed from the Droop compliant sets from the Example 1 ballots. For each of these lists, the top N candidates form a Droop compliant set of N winners for any N.

Due to the necessity to produce a party list of candidates for the European parliament elections in 1999, some political party leaders in the United Kingdom considered conducting a ranked candidate ballot election of their members to produce a candidate list (Rosenstiel, C 1998; Otten, J. 1998). Conventional STV does not always comply with house monotonicity so it cannot, in general be used to produce a candidate list. For example, the winner of the conventional STV one-winner election for the Example 1 ballot set is C and the winners of the three-winners election are A, B and D. (Since compliance with house monotonicity is a necessary but not sufficient condition for compliance with coherence, this means that STV also fails coherence. This is shown explicitly in Section IV). Proposals to modify STV to produce a house monotone candidate list are either bottom-up or top-down, as described below.

Bottom-up methods produce a candidate list by determining the candidate in the lowest position, then the second lowest, etc. They determine the $N^{th}$ position candidate using a method to produce a set of N-1 winners from N candidates in which all candidates in positions N+1 and lower have been excluded from the ballots. The loser of the N-1-winners election is placed in the $N^{th}$ position.

A bottom-up method for conventional STV is described by Rosenstiel (Rosenstiel, C 1998). Rosenstiel's method applied to the Example 1 ballots produces the list A>D>B>C.

Aziz, et al. (Aziz et al. 2025) have shown that every bottom-up method based on a Droop proportional method is house monotonic and Droop proportional. This is because if more than $KQ_N$ ballots prefer $L$ candidates above all others, then more than $KQ_{N+1}$ ballots prefer at least the same $L$ candidates above all others. This is because $KQ_N > KQ_{N+1}$ and every subset of a solid coalition is itself a solid coalition preferring, at least, the same preferred candidates as the solid coalition it is a subset of. Therefore, there will always be a Droop compliant set for N winners that is a subset of one of the Droop compliant sets for N+1 winners.

The highest ranked candidate of a bottom-up candidate list can be different from the winner of the one-winner method that the bottom-up method is based upon.

This is the case for bottom-up STV applied to the Example 1 ballots. The top position candidate using bottom-up STV is A, while the winner of conventional STV for one winner is C.

A top-down method avoids this discrepancy by producing a candidate list by determining the candidate in the highest position, then the second highest, etc. It determines the $N^{th}$ position candidate by using a method to produce a set of N winners that has been modified to guarantee the election of the first N-1 candidates in the list. The winner of the N-winners election that is not one of the candidates in positions 1 through N-1 is placed in the $N^{th}$ position.

The highest ranked candidate of a top-down candidate list is, by construction, the same as the winner of the one-winner method that the top-down method is based upon.

A top-down method based on conventional STV, in which previously elected candidates are protected from exclusion, and, after the election of the first not-previously elected candidate, all previously elected candidates are reelected, and the count ends, was proposed by Otten for the the Green Party of England and Wales (Otten, J. 1998). It does not always comply with Droop proportionality. This can be seen by applying Otten's method to the Example 2 ballot set, in which Candidate E is a clone of Candidate B.

Example 2 ballot set: 60:A>D>C>B>E, 26:B>E>C>D>A, 25:E>B>C>D>A, 45:C>A>D>B>E, 44:D>C>A>B>E.

The Otten winners in the three-winners election are A, C, and D. However, Droop proportionality requires that one of the winners be from the 51-ballot solid coalition preferring candidates B and E.

Since house monotonicity and coherence are properties of divisor methods, it is natural to explore adapting divisor methods to ranked candidate voting to achieve these properties. Aziz, et al. (Aziz et al. 2025) describe a bottom-up divisor method for ranked candidate ballots based on Phragmén's procedure for approval-style ballots, that we discuss and show compliance with house monotonicity, coherence, and Droop proportionality in Section IV. The method is not free riding resistant.

Aziz et al. (Aziz et al. 2025) developed Solid Coalition Refinement, a top-down house monotonic Droop compliant method not based on Phragmén's procedure. This method fails proportional candidate sharing.

**Section IV: Approval, Quota, and Bottom-up Phragmén**

In 1895, the mathematician Edvard Phragmén (Phragmén, E. 1895; 1899) described a divisor election method for approval style ballots that satisfies house monotonicity, coherence, and Droop proportionality.

Phragmén assigned priorities to candidates to determine which to elect next. The priority of candidate *C* is $P_C = V_C/(1+S_C)$, where $V_C$ is the number of ballots approving the election of candidate C and $S_C$ is the portion of winning candidate seats the ballots approving C have already contributed to electing. Candidate C is elected if it has the highest priority. If candidate C is elected, the seat load of each ballot that contributed to electing C is updated to $(S_C+1)/V_C$, the inverse of the priority.

A little over a century later, Olli Salmi (Salmi, Olli 2002) and Douglas Woodall (Woodall, Douglas R. 2003) developed ranked-candidate election methods based on Phragmén's method.

To determine priorities, their methods assign support from a ballot to its highest ranked candidate that is not elected or excluded. They borrow, from STV, the idea of using a quota to determine if an election or exclusion of a candidate should occur, electing the highest priority candidate when its priority exceeds the quota and excluding the lowest priority candidate when no candidate's priority exceeds the quota. These methods, like STV, are compliant with Droop proportionality (Woodall, Douglas R. 2003), but because of their reliance on a quota they violate house monotonicity and coherence as demonstrated by the examples below.

Quota-based Phragmén applied to the Example 1 ballot set violates house monotonicity. The winner of the quota-based Phragmén one-winner election for the Example 1 ballots set is C and the winners of the three-winners election are A, B and D, the same house-monotonicity violating results obtained by conventional STV.

Quota-based Phragmén applied to the Example 3 ballot set, which is the Example 1 ballot set with the addition of 45 ballots for candidate E, violates coherence.

Example 3 ballot set: 60:A>D>C>B, 51:B>C>D>A, 45:C>A>D>B, 44:D>C>A>B, 45:E

The winners of the three-winners conventional STV and quota-based Phragmén elections for the Example 3 ballot set, are A, B, and C, whereas the winners for the Example 1 set, in the absence of the ballots for candidate E, are A, B, and D. The violation of coherence comes about because the quota is determined by size of the entire ballot set and changes when the 45 E votes are added, effecting when to elect and when to exclude non-E candidates.

However, there is an important case where the methods of Salmi and Woodall do not rely on a quota. When M-1 winners are required to be elected from the remaining M candidates, the method elects M-1 candidates, one after the other, without excluding any candidates and so the quota is not relied on for those cases.

To see this, note that after the election of P candidates, the average of the M-P remaining candidate priorities weighted by their denominators is $\frac{\sum(S_C+1)V_C/(S_C+1)}{\sum(S_C+1)} = \frac{V_{Tot}}{M} = Q_{M-1}$. The highest the remaining M-P priorities is always above average. So the highest priority exceeds $Q_{M-1}$, and so the highest priority candidate will be elected.

Electing M-1 winners from M candidates is exactly what is required in bottom-up methods for producing a candidate list. Therefore, since quota-based Phragmén is Droop compliant, and, Aziz, et al. (Aziz et al. 2025) proved that all bottom-up methods relying on a Droop compliant method are themselves Droop compliant, so is bottom-up Phragmén. It is house monotonic by construction, and it is coherent because no quotas are used.

The proportional list produced by bottom-up Phragmén, A>D>B>C, is the same list produced by bottom-up STV.

**Section V: The Top-down Phragmén Method**

In this section we present a top-down ranked-candidate election method based on Phragmén's method for approval ballots that does not rely on a quota to determine when to elect and when to exclude. We show in Section VI that the method is house-monotone, coherent, and Droop proportional. In Section VII we show that the method is free riding resistant, unlike STV and bottom-up Phragmén, and that it, along with STV and bottom-up Phragmén, complies with shared candidate proportionality.

For the M-winner election we label the M-1 winners of the M-1-winner election as *previously elected* and all other candidates as *hopeful*.

To adhere to both top-down house monotonicity and Droop proportionality, Phragmén priorities cannot be calculated in the normal unconstrained way. To see why this is, consider the 3-winner election using ballots from Example 1, in which candidates C and A are previously elected into position 1 and 2 and we are determining the third place winner. Following the normal Phragmén procedure, after the election of candidate A, hopeful candidate D has Phragmén priority 52, which is higher than hopeful candidate B's Phragmén priority 51, even though {A, B, C} is a Droop compliant set while {A, C, D) is not. The unconstrained Phragmén priority for D incorrectly exceeds the priority for B because it includes 44 ballots ranking D first and previously elected candidate C second but does not include the seat load from C or A. D is elected as a free rider.

We prevent this from happening by properly accounting for the seat loads of previously elected candidates even if they are less preferred than hopeful candidates.

The procedure requires that we create an approval ballot from each ranked ballot that approves the highest ranked hopeful candidate and a set of previously elected candidates preferred over other previously elected candidates. When calculating the priorities of hopeful candidates, the method treats the same hopeful candidates on ballots that approve different sets of previously elected candidates as if they are different hopeful candidates. We do this by appending to the hopeful candidate a number whose binary representation uniquely identifies a set of previously elected candidates. The previously elected candidate that is in $N^{th}$ place on the overall winning candidate list is assigned the number $2^{N-1}$ and the sum of these approved candidate numbers is appended to the approved hopeful candidate. The approval ballot is designated by its hopeful candidate and the appended number. For example, if the previously elected winners are 1:A, 2:B, 3:C, 4:D, and a ranked ballot ranks candidates C>B>E>A>F>D, in which candidates E and F are hopeful, then an approval ballot that approves hopeful Candide E and previously elected candidates B and C is designated E6. If there are no previously elected candidates ranked higher than the highest ranked hopeful candidate, for example E>C>B>A>F>D, the associated approval ballot does not initially approve any previously elected candidates and is designated E0. Later, in step 6 of the procedure, the the number of approved previously elected candidates on a ballot can increase to include those that are less preferred than the highest ranked hopeful candidate, even less preferred than lesser preferred hopeful candidates. For example, an approval ballot associated with the ranked ballot C>B>E>A>F>D can approve hopeful candidate E and all previously elected candidates, including A and D. In that case its designation is E15.

The details of the procedure are as follows: For the M-winner election, label the M-1 winners of the M-1-winner election as "previously elected" and all other candidates as "hopeful." Set all seat values to zero and perform the following steps:

1. Create an approval ballot associated with each ranked ballot. If there is at least one ranked hopeful candidate then, initially, the approval ballot approves the most preferred hopeful candidate and all previously elected candidates that are preferred to the highest ranked hopeful candidate. If there are no ranked hopeful candidates, the approval ballot approves all ranked previously elected hopeful candidates. In general, each approval ballot approves at most one hopeful candidate and can approve any number of previously elected candidates. If there is an approved hopeful candidate on the approval ballot, the ballot is designated with its approved hopeful candidate and its approved previously elected candidate number.

2. Perform the Phragmén procedure on the approval ballots for approved previously elected candidates only, to reelect all previously elected candidates, sequentially electing the previously elected candidate with highest priority and updating the seat loads according to the Phragmén procedure until all previously elected candidates have been reelected. Hopeful candidates are not considered approved at this stage. All ballots that approve a previously elected candidate contribute to that candidate's priority. Tiebreaker: Ties in the priority of previously elected candidates for the M-winner election can be broken with the priorities from the nearest previous round that breaks the tie. This is done for the 3-winner election in the example in the next section. If there is no previous round of the M-winner election that breaks the tie, the tie can be broken using the winning candidate list order.

3. Calculate the Phragmén priority for hopeful candidates with different designations separately. For example, every approval ballot designated E0, E6, or E15 is included in the calculation for the priority of E0, E6, and E15 as if they were different candidates.

4. The priority of each hopeful candidate is the maximum priority of the solid coalition priorities that approve that hopeful candidate. For example if all ballots approving E were designated either E0, E6, or E15 then E's priority is the maximum of the priorities of E0, E6, and E15. In steps 5 and 6, we call the lowest priority hopeful candidate L. If all of the ballots that approve L have the same designation, for example they are all L6 and there are no L ballots with any other designation, then go to step 5. Otherwise go to step 6. There are more options to break ties in the priorities of hopeful candidates for the M-winner election because each hopeful candidate can have several different designations, each with its own priority. Tiebreaker: Ties can be broken by considering priorities in addition to the largest for each hopeful candidate. If that does not break the tie, ties can be broken by considering the priorities from the nearest previous round that breaks the tie. If there is no previous round of the M-winner election that breaks the tie, the tie can be broken using the exclusion order for the M-1-winner election. If the tie occurs in the 1-winner election and there is no previous round that breaks the tie, the tie must be broken using an initial order of the candidates, chosen at random or through other means.

5. Exclude candidate L from the ranked ballot set. If only one hopeful candidate remains, elect that candidate to position M and end the count. Otherwise, return to step 1 with L excluded. Each time this step is performed an additional candidate is excluded from the original ballot set.

6. Consider only those ballots that approve L and the smallest number of previously elected candidates. Keep in mind that there are, in general, many designations with the same number of approved candidates and they are, in general, not in consecutive numerical order. For example, if the previously elected candidates are 1:A, 2:B, 3:C then L0 approves no previously elected candidates. L1, L2, and L4 all approve 1 previously elected candidate. L3, L5, and L6 approve 2 previously elected candidates, and L7 approves 3. On every approval ballot under consideration for which there is at least one previously elected candidate on its associated ranked ballot that has not yet been approved, approve the most preferred of these and update the designation to include it. For example, if the smallest number of approved previously elected candidates is 1, and amongst those is the ballot C>L>D>E>A>B>C, where D and E are other hopeful candidates, then that L4 ballot, which approves L and C, will now also approve A and be redesigned L5. On every approval ballot under consideration that approves every previously elected candidate ranked on its associated ranked, approve all M-1 previously elected candidates and update the designation to $2^{M-1}$-1, which is the designation for including all M-1 previously elected candidates. For example, an L4 ballot ranking C>L>D, with E, A, B, and D unranked, will now approve all previously elected candidates A, B, C and D, and be redesigned L7. Unelect all previously elected candidates, set seat loads to zero, and go to step 3 to elect all previously elected candidates again.

The approval ballot associated with each ranked ballot can be represented as an approval cutoff added to each ranked ballot. All previously elected candidates ranked above the approval cutoff and the most preferred hopeful candidate ranked above the cutoff are approved, with all other candidates unapproved. Initially the cut off is placed to approve only the most preferred hopeful candidate and those previously elected candidates more preferred than the most preferred hopeful candidate. The cutoff is lowered to include an additional ranked previously elected candidate if there is one when step 6 is invoked.

If a ranked ballot is not fully ranked, all unranked previously elected candidates are interpreted as ranked equally immediately below the lowest ranked candidate, which is why if there are no previously elected candidates ranked below the cutoff, the effect of lowering the approval cutoff is that all previously elected candidates are approved, and the designation of the approved hopeful candidate is increased to $2^{M-1}$-1. An alternative is to proportionally complete unranked ballots. Some form of ballot competition is necessary to guarantee that before a candidate is excluded, all ballots in the solid coalition ranking that candidate the highest hopeful candidate are given the opportunity to contribute to that candidate's priority.

The motivation for the method is to adjust the approval cutoff of ballots minimally until a configuration is found where the lowest priority candidate cannot remove itself from lowest priority position by further downward adjustment of the approval cutoff on ballots that prefer that hopeful candidate over others. The condition in step 4 to exclude the lowest priority candidate when there is only one designation for that candidate, rather than the natural endpoint of applying step 6, which would be to exclude the lowest priority candidate when there is only one designation for that candidate *and* it includes all previously elected candidates, is that when there is only one designation, approving additional candidates on those ballots can only decrease its priority and only increase the priorities of other candidates. Such a candidate is fated to remain the lowest priority candidate and to be excluded because its sole designation would eventually approve all previously elected candidates after repeated application of step 6.

The procedure creates approval ballots that can approve previously elected candidates that are not only less preferred than the highest ranked hopeful candidates, but also less preferred than lower ranked hopeful candidates, and even not ranked at all. This is required for free rider resistance.

**Section VI: A Top-Down Phragmén Example**

Top-down Phragmén is demonstrated on the Example 1 ballot set, for one-winner, two-winners, and three-winners elections, in Tables 1, 2, and 3.

Table 1: One Winner

| | A0 | B0 | C0 | D0 | Actions |
|---|---|---|---|---|---|
| 1 | 60 | 51 | 45 | 44 | Exclude D |
| 2 | 60 | 51 | 89 | X | Exclude B |
| 3 | 60 | X | 140 | X | Exclude A. C is elected. |

Table 1 shows the votes for candidates at each step of the one-winner election. The top-down method for one winner is the same as Instant Runoff because there are no previously elected candidates, therefore the associated approval ballot for each ranked ballot just approves the highest ranked candidate and priorities are the same as votes. Candidates, D, B, and A are excluded because they have the fewest votes at each round. Candidate C, the last remaining hopeful candidate, is elected to the first position on the proportional list.

Table 2: Two winners, C is previously elected.

| | C | A0 | A1 {C} | B0 | D0 | Actions |
|---|---|---|---|---|---|---|
| 1 | 45 | 60 | ~~22.5~~ | 51 | 44 | Elect C, Exclude D |
| 2 | 89 | 60 | ~~44.5~~ | 51 | X | Elect C, Exclude B. Elect A |

Table 2 shows the priorities for candidates at each step of the two-winners election. Each round begins with the reelection of the previously elected candidate C. The priorities for designated ballot sets are shown after C’s reelection. If there is more than one designation for a hopeful candidate, the priorities that are not the largest are struck through. In round 1, the associated approval ballots with C previously elected are 60 A0, 51 B0, 45 C>A1, 44 D0. A is the only candidate with more than one designation. After C’s reelection, A0 has a larger priority than A1. So A ’s priority is A0’s. All of the other hopeful candidates have their 0 designated priority as their one and only priority. D has the lowest priority and is excluded. In round two, the associated approval ballots are 60 A0, 51 B0, 89 C>A1. After C’s election, A0 again has a larger priority than A1. B has the lowest priority and is excluded. A is elected to the second position on the proportional list.

Table 3: Three winners, A and C are previously elected.

| | A | C | B0 | D0 | D1 {C} | D2 {A} | D3 {C,A} | Actions |
|---|---|---|---|---|---|---|---|---|
| 1 | 105 | 31.5 | 51 | 44 | ~~0~~ | ~~38.18~~ | ~~18.53~~ | Elect A, Elect C, redesignate D0 |
| 2 | 105 | 62.3 | 51 | ~~0~~ | ~~25.79~~ | 38.18 | ~~26.13~~ | ElectA, Elect C, redesignate D1 and D2 |
| 3 | 149 | 74.5 | 51 | ~~0~~ | ~~0~~ | ~~0~~ | 49.67 | Elect C, Elect A, Exclude D, Elect B |

Table 3 shows the priorities for candidates at each step of the three-winners election. Each round begins with the reelection of the previously elected candidates, A and C, in order of priority. In the first two rounds A has the higher priority and is reelected before C. In the third round they are tied. A is elected first according to the tiebreaker. The priorities of the previously elected candidates shown in the table are their priorities at the time of their reelection. The priorities of the designated hopeful candidates shown in the table are those after the reelection of all previously elected candidates. If there is more than one designation for a hopeful candidate, the priorities that are not the largest are struck through. In round 1, the approval ballots are 60 A>D2, 51 B0, 45 {C,A}> D3, 44 D0. After the reelection of the previously elected candidates. D's highest priority D0 is the lowest of the hopeful candidates. This is the first occurrence in this example when the lowest priority candidate has more than one designation so D undergoes redesignation instead of exclusion. D0 is the designation with the fewest approved candidates. Its ballots now approve C and are designated D1. In round 2 the approval ballots are 60 A>D2, 51 B0, 45 {A,C}> D3, 44 C> D1. D still has the lowest priority. D is again undergoes redesignation. D1 and D2 have age fewest approved candidates. These ballots now approve A and C and are redesignated D3. In round 3, the approval ballots are 149 {A,C}> D3, 51 B1. D again has the lowest priority and is excluded. B is elected to the third position, leaving D to be elected to the fourth position.

The proportional list produced by top-down Phragmén is C>A>B>D, consistent with Droop proportionality.

**Section VII: House monotonicity, Coherence, and Droop Proportionality**

In this section we show that top-down Phragmén is house monotone, coherent, and Droop proportional. Consideration is restricted, for simplicity, to ballot sets for which there are no ties for exclusions.

**Claim 1: House Monotonicity**. The winners of the M-winners election always include the M-1 previously elected candidates. **Proof:** The M-winner election always reelects the M-1 previously elected candidates and never excludes then.

**Claim 2**: **Coherence**. If there are two sets of candidates, and each ballot *completely* ranks one of the two sets of candidates without ranking any of the other set, the relative order of the candidates in each set are unchanged from the order obtained by counting ballots ranking each set separately. **Proof:** We call the two sets of candidates A candidates and B candidates. The approval ballots associated with the A (B) ranked ballots approve only A (B) candidates so the priorities of A (B) candidates depend only on A (B) ballots. Therefore the reelection, election, and exclusion, of an A (B) candidate will not impact B (A) ballots in any way. Redesignation requires a little more consideration. If L from the A (B) candidates is the hopeful candidate with smallest priority and L has more than one designation, then each A (B) ballots that gets redesignated will have an additional previously elected A (B) candidate added to the approved ballot, as a consequence of the requirement that each A (B) ballot entirely ranks all A (B) candidates. B (A) ballots will not be impacted in any way. The proof does not apply and should not apply if the A (B) candidates are incompletely ranked on some ballots. In that case, counting the A (B) ballots separately has a different meaning than counting them together because incomplete A (B) ballots get completed with only unranked A (B) candidates when counted separately but with all unranked candidates when counted together.

For claims 3-7, we assume that every ballot ranks all candidates so there is no ambiguity concerning any ballot's candidate preferences. We also define the *final stage* of the M-winners count, it is the final round of the procedure one would follow if one continued the M-winners procedure with the one remaining hopeful candidate. This is necessary because the proofs for claims 3-7 require us to know the priority of a candidate after the exclusion of the last candidate in the previous stage.

**Claim 3**: In the final stage of any M-winners count, the priority of the new M-winner exceeds $Q_M$. **Proof:** The last remaining hopeful candidate, at the final stage when the new M-winner is the only non-excluded hopeful candidate and all ballots that rank that candidate are in the same designation, has the support of all of the voters for fully ranked ballots. Its priority is $V_{Tot}/(P+1)$ where P is the number of approved previously elected candidates in the single remaining designation for the new M-winner. This number is less than or equal to M-1 so the priority is greater than or equal to $V_{Tot}/M > V_{Tot}/(M+1) = Q_M$. Therefore, the priority of the new M-winner exceeds $Q_M$ in the final stage of the count.

**Claim 4:** For every $N \geq M$, there is a stage in every M-winners count when the priority of the eventual new M-winner exceeds $Q_N$ for the first time and no redesignated or excluded candidate up to that stage had a priority exceeding $Q_N$. **Proof**: The claim is satisfied if at the first stage of the M-winners count, the priority of the new M-winner exceeds $Q_N$. If the priority of the new M-winner does not exceed $Q_N$ at the first stage of the M-winners count, the candidate that is to be redesignated or excluded, which has the lowest priority of all hopeful candidates, must also have a priority that does not exceed $Q_N$. Then, if the new M-winner has a priority exceeding $Q_N$ in the second stage, the claim is satisfied. If the new M-winner does not have a priority exceeding $Q_N$, then, as with stage 1, the hopeful candidate that is to be redesignated or excluded has the lowest priority and therefore also not exceeding $Q_N$. This is repeated until the stage when the priority for the new M-winner exceeds $Q_N$ for the first time. There will always be such a stage because the new M-winner is never excluded and at the final stage, the votes for the new M-winner exceeds $Q_M$ which exceeds $Q_N$ for all $N \geq M$.

**Claim 5:** The only way that the new M-winner could not be in any N-winner Droop compliant set is if the ballots can be divided into non-intersecting solid coalitions such that the i[th] solid coalition contains more than $K_iQ_N$ ballots with $\sum_i K_i = N$, and 2) each solid coalition must prefer at least $K_i$ candidates above all others, 3) each set of candidates is completely non-intersecting with the other sets, with no candidates in common, and 4) none of the at least Ki candidates from each set are the new M-winner. **Proof:** By the requirements of Droop proportionality, N candidates composed of $K_i$ preferred candidates from each solid coalition are the only N-winner Droop compliant sets for this ballot configuration. By the requirement of the claim, the new M-winner is not in any of them. If one changes the requirements of the claim, in particular, if the i[th] solid coalition prefers fewer than $K_i$ candidates above all others, or if there are any shared candidates amongst the preferred candidate sets of the solid coalitions, then the number of candidates that must be in the winning set for Droop compliance is fewer than N, allowing the new M-winner to be one of the N winners without violating Droop compliance.

**Claim 6:** If there was a new M-winner-excluding ballot set, as described in Claim 5, then when applying the M-winner procedure to such a ballot set, the new M-winner could not have a priority that exceeds $Q_N$ before a candidate is redesignated or excluded with a priority that exceeds $Q_N$. **Proof:** The new M-winner initially could not have a priority that exceeds $Q_N$ because it is not the highest ranked hopeful candidate on any of the ballots in the solid coalitions that collectively prefer at least N candidates over the new M-winner. So the contribution to the new M-winner's priority from those ballots is zero. There could be a group of ballots, fewer than $Q_N$ in number, that are not in any of the aforementioned solid coalitions, but even if the new M-winner was the highest ranked hopeful candidate on all of those ballots, its priority, equal to the number of those ballots if those ballots approved no previously elected candidates, and lower otherwise, could not exceed $Q_N$. The priority of the new M-winner can only exceed $Q_N$ once every hopeful candidate preferred to the new M-winner in one of the solid coalitions, in what follows the i[th], has been excluded, and even then, only if the number, P, of previously elected candidates approved by the ballots of the i[th] solid coalition is fewer than $K_i$, for then the the new M-winner will be the highest ranked hopeful candidate in the ballot of that solid coalition and have a priority from that solid coalition's designation greater than $K_iQ_N/P > Q_N$. However, for this to happen, the last excluded candidate from this coalition would have been top ranked on those ballots before exclusion and would therefore have also had a priority exceeding $Q_N$, by the same reasoning, before one of its redesignations or its exclusion.

**Claim 7: Droop Proportionality of the new M-winner**. The new M-winner is in at least one Droop compliant set of N winners for all $N \geq M$. **Proof:** As shown in the proof for Claim 4, when following the M-winner procedure, the priority of the new M-winner exceeds $Q_N$ for $N \geq M$ for the first time at a step when the priority of all previous redesignated or excluded candidates did not exceed $Q_N$. As shown in the proof of Claim 6, if there was a ballot set in which the new M-winner was not in any N-winners Droop compliant set, then when applying the M-winners procedure on that ballot set, The new M-winner's priority could only exceed $Q_N$ after another candidate had a priority that exceeded $Q_N$ before at least one of its redesignations or its exclusion. Therefore, there is at least one N-winners Droop compliant set that includes the new M-winner for all $N \geq M$. We call this property of the new M-winner **core support** because the M-winner has a core support of voters who stick with them as the number of winning options increase.

For M =1, the M-winners procedure is the IRV procedure. Therefore, the IRV winner is in at least one Droop compliant set of N winners for all N and therefore complies with core support. This is an interesting property of the IRV winner. The Condorcet winner is not guaranteed to be in at least one Droop compliant set of N winners for all N, except for N = 1 (Woodall, Douglas R. 1994).

The argument in Claims 3-7 also apply to winners of the M-winners election using the STV procedure, where "previously elected candidates" refer to candidates that the STV procedure elects before the winner under consideration, and we just have exclusions instead of exclusions and redesignations. Therefore, winners of the STV procedure for M-winners are also in at least one Droop compliant set for N winners for all $N \geq M$. However, there is no guarantee that STV selects one of these sets for the N-winners election. This is how STV winners satisfy core support although STV is not house monotonic.

**Claim 8: Droop proportionality of the M-winners.** The M-winners are, together, in at least one Droop compliant set of N winners for all $N \geq M$. **Proof:** This is an inductive proof. We will show that if all M-1 winners of the M-1-winners election are, together, in at least one Droop compliant set of N winners for all $N \geq M$, as is the case when M-1=1, because the IRV winner satisfies this requirement, then it is also true for the M-winners. If it were true for the M-1-winners, then the new M-winner excluding ballot set in Claim 5, must include all of the M-1 previously elected winners amongst its preferred candidates for there to be a ballot set that has at least one Droop compliant set of N-winners containing all the M-1-winners but none containing all the M-winners. But this does not alter the proofs of Claim 6 and 7, so such a ballot set is impossible and there will always be at least one Droop compliant set of N winners containing all M winners from the M-winners election. Each of the M-winners have have various levels of **broad support** in that they are all in at least one Droop compliant set of N winners for N=M and are also in at least one Droop compliant set for N<M for N as small as J for the candidate who is in the $J^{th}$ position in the winning candidate list, as voters are drawn towards them as the number of winning possibilities decease from M to J, indicating an inclusive appeal.

**Section VIII: Free Rider Resistance and Shared Candidate Proportionality**

**Free Rider Resistance:** A free rider is someone who ranks a popular candidate lower or does not rank it so their ballot will retain more strength for other candidates. Consider the 2-winner election for the ballot set 25: A>B>C, 24: B > A > C, 51: A>C>B. STV methods as well as quota based and bottom-up Phragmén elect A and B. B is elected because the 24 ballots ranking B first contribute to B's election without having to contribute to A's election. Top-down Phragmén elects A and C because it recognizes that the 24 B>A>C voters do not have enough votes by themselves to elect B. When combining with the 25 A>B>C voters to elect B from a solid coalition preferring A and B, those 24 ballots are associated with approval ballots that approve A and B and contribute to A's election just as much as if they had ranked A first and B second. The result is the same if the 24 voters voted, B>A>C, B>C>A, or B with no second or third candidates ranked.

Free rider resistance is incompatible with the later-no-harm criterion complied with by STV. However there is a weaker version of later-no-harm that top-down Phragmén does comply with. If M candidates are in position 1 to M in the candidate list, changes to ballots that occur below all M candidates cannot change the 1st through Mth winners in any way. This is so because the priority of a hopeful candidate is unaffected if there are changes made to ballots that occur only below the hopeful candidate *and* below all previously elected candidates because these changes do not change the associated approval ballots that the count actually uses. Because of the *and*, this is a more restrictive form of later-no-harm than that obeyed by STV.

**Proportional Candidate Sharing**: If a ballot set is changed by adding a new candidate and a new highest rank to each ballot in which every voter places the new candidate in the new highest rank, leaving all other rankings otherwise unchanged, then top-down Phragmén applied to the new ballot set produces a winning candidate list composed of the the new candidate at the top with the original winning candidate list below the new candidate. STV also adheres to this property. But there are voting methods, such as Aziz et al's Solid Coalition Refinement, that do not. Consider the three-winner election for the ballot set 50 A>B>C, 30 A>D>E. top-down, quota and bottom-up Phragmén along with STV all elect A, B, and D. For Phragmén methods, after the election of A and B, D's priority is 30/(3/8+1) =21.82, which exceeds C's priority 50/(5/8+2)=19.05. But methods that determine the priority of a solid coalition by dividing its number of ballots by the number of preferred candidates, without regard for how previously elected candidates are shared with other solid coalitions, elect A, B and C. For these methods, after A and B are elected the priority to elect C is 50/3=16.67 which exceeds the priority to elect D, which is 30/2=15. Adding a universally preferred candidate changes the Phragmén priority to elect candidate B from $V_B/(S_B+1)$ to $V_B/(V_B/V_{Tot} + S_B+1)$, where $V_B/V_{Tot}$ is the proportional share of the B ballots contribution towards electing candidate A. The inverse of this priority is $(V_B/V_{Tot} + S_B+1)/V_B = 1/V_{Tot} + (S_B+1)/V_B$, which is just the inverse of the original priority plus a constant shift. A constant shift applied to all priorities does not change their relative order. Therefore, following the election of the universally preferred candidate to first place in the candidate list, the original candidates will be elected in their original order below it.

**Section IX: Summary**

Top-down Phragmén, a ranked-candidate election method, is presented that produces a house-monotone, coherent, and Droop proportional candidate list that is free riding resistant and complies with proportional candidate sharing. The highest ranked candidate is the IRV winner, which is in at least one Droop compliant set of N winners for all N. Table 4 compares Top-down Phragmén to other Droop compliant proportional methods. It shares house monotonicity with bottom-up Phragmén, bottom-up STV and Aziz et al's Solid Coalition Refinement (SCR), which are the other methods that do not use a quota to decide when to elect and when to exclude a candidate. It shares coherence with SCR and bottom-up Phragmén, the other methods that do not use a quota at all (the decision to elect or exclude and how much voting power to transfer to lower ranked candidates). It shares free rider resistance with SCR, the other top-down method. It shares proportional candidate sharing with the other Phragmén and STV methods.

Table 4: Properties of Proportional Ranked Candidate Voting Methods

| | **Droop Proportionality** | **House Monotonicity** | **Coherence** | **Free Rider Resistance** | **Proportional Candidate Sharing** |
|---|---|---|---|---|---|
| **Top-down Phragmén** | Yes | Yes | Yes | Yes | Yes |
| **SCR** | Yes | Yes | Yes | Yes | No |
| **Bottom-up Phragmén** | Yes | Yes | Yes | No | Yes |
| **Bottom-up STV** | Yes | Yes | No | No | Yes |
| **Quota based Phragmén and STV** | Yes | No | No | No | Yes |